\documentclass[twocolumn,floatfix,showpacs,aps,prl,superscriptaddress]{revtex4}

\usepackage{graphicx}
\usepackage{dcolumn}
\usepackage{amsmath}
\usepackage{units}

\RequirePackage{xspace}

\usepackage{relsize}
\def\babar{\mbox{\slshape B\kern-0.1em{\smaller A}\kern-0.1em
    B\kern-0.1em{\smaller A\kern-0.2em R}}}

\def\Kbar  {\kern 0.2em\overline{\kern -0.2em K}{}\xspace}

\def\Kz    {\ensuremath{K^0}\xspace}
\def\Kzb   {\ensuremath{\Kbar^0}\xspace}
\def\KzKzb {\ensuremath{\Kz \kern -0.16em \Kzb}\xspace}
\def\Kp    {\ensuremath{K^+}\xspace}
\def\Km    {\ensuremath{K^-}\xspace}

\def\KpKm  {\ensuremath{\Kp \kern -0.16em \Km}\xspace}

\def\Dbar    {\kern 0.2em\overline{\kern -0.2em D}{}\xspace}

\def\Dz      {\ensuremath{D^0}\xspace}
\def\Dzb     {\ensuremath{\Dbar^0}\xspace}
\def\DzDzb   {\ensuremath{\Dz {\kern -0.16em \Dzb}}\xspace}
\def\Dp      {\ensuremath{D^+}\xspace}
\def\Dm      {\ensuremath{D^-}\xspace}

\def\DpDm    {\ensuremath{\Dp {\kern -0.16em \Dm}}\xspace}

\def\Bbar    {\kern 0.18em\overline{\kern -0.18em B}{}\xspace}

\def\Bz      {\ensuremath{B^0}\xspace}
\def\Bzb     {\ensuremath{\Bbar^0}\xspace}
\def\BzBzb   {\ensuremath{\Bz {\kern -0.16em \Bzb}}\xspace}
\def\Bu      {\ensuremath{B^+}\xspace}
\def\Bub     {\ensuremath{B^-}\xspace}

\def\BpBm    {\ensuremath{\Bu {\kern -0.16em \Bub}}\xspace}

\def\BorBbar    {\kern 0.18em\optbar{\kern -0.18em B}{}\xspace}
\def\DorDbar    {\kern 0.18em\optbar{\kern -0.18em D}{}\xspace}
\def\KorKbar    {\kern 0.18em\optbar{\kern -0.18em K}{}\xspace}

\mathchardef\Upsilon="7107
\def\Y#1S{\ensuremath{\Upsilon{(#1S)}}\xspace}

\mathchardef\Deltares="7101
\mathchardef\Xi="7104
\mathchardef\Lambda="7103
\mathchardef\Sigma="7106
\mathchardef\Omega="710A

\def\Deltabar{\kern 0.25em\overline{\kern -0.25em \Deltares}{}\xspace}
\def\Lbar{\kern 0.2em\overline{\kern -0.2em\Lambda\kern 0.05em}\kern-0.05em{}\xspace}
\def\Sigbar{\kern 0.2em\overline{\kern -0.2em \Sigma}{}\xspace}
\def\Xibar{\kern 0.2em\overline{\kern -0.2em \Xi}{}\xspace}
\def\Obar{\kern 0.2em\overline{\kern -0.2em \Omega}{}\xspace}
\def\Nbar{\kern 0.2em\overline{\kern -0.2em N}{}\xspace}
\def\Xb{\kern 0.2em\overline{\kern -0.2em X}{}\xspace}

\newcommand{\tev}{\ensuremath{\mathrm{\,Te\kern -0.1em V}}\xspace}
\newcommand{\gev}{\ensuremath{\mathrm{\,Ge\kern -0.1em V}}\xspace}
\newcommand{\mev}{\ensuremath{\mathrm{\,Me\kern -0.1em V}}\xspace}
\newcommand{\kev}{\ensuremath{\mathrm{\,ke\kern -0.1em V}}\xspace}
\newcommand{\ev}{\ensuremath{\mathrm{\,e\kern -0.1em V}}\xspace}
\newcommand{\gevc}{\ensuremath{{\mathrm{\,Ge\kern -0.1em V\!/}c}}\xspace}
\newcommand{\mevc}{\ensuremath{{\mathrm{\,Me\kern -0.1em V\!/}c}}\xspace}
\newcommand{\gevcc}{\ensuremath{{\mathrm{\,Ge\kern -0.1em V\!/}c^2}}\xspace}
\newcommand{\mevcc}{\ensuremath{{\mathrm{\,Me\kern -0.1em V\!/}c^2}}\xspace}

\def\mus  {\ensuremath{\rm \,\mus}\xspace}

\def\mus        {\ensuremath{\,\mu{\rm s}}\xspace}    

\def\pep2{PEP-II}

\def\gsim{{~\raise.15em\hbox{$>$}\kern-.85em
          \lower.35em\hbox{$\sim$}~}\xspace}
\def\lsim{{~\raise.15em\hbox{$<$}\kern-.85em
          \lower.35em\hbox{$\sim$}~}\xspace}

\xspace

\def\jetset74   {\mbox{\tt Jetset \hspace{-0.5em}7.\hspace{-0.2em}4}\xspace}

\newcommand{\BABARPubYear}    {07}
\newcommand{\BABARPubNumber}  {x}
\newcommand{\SLACPubNumber} {x}

\def\figurebox#1#2#3{%
    \def\arg{#3}%
    \ifx\arg\empty
    {\hfill\vbox{\hsize#2\hrule\hbox to #2{\vrule\hfill\vbox to #1{\hsize#2\vfill}\vrule}\hrule}\hfill}%
    \else
    {\hfill\epsfbox{#3}\hfill}%
    \fi}

\long\def\inst#1{\par\nobreak\kern 4pt\nobreak
    {\it #1}\par\vskip 10pt plus 3pt minus 3pt}

\begin{document}

\preprint{\babar-PUB-\BABARPubYear/\BABARPubNumber}
\preprint{SLAC-PUB-\SLACPubNumber}

\begin{flushleft}
 BABAR-PUB-07/040;  \\
 SLAC-PUB-12709; \\
 arXiv:0708.1303 [hep-ex] \\
 Phys. Rev. Lett. {\bf 99}, 201801 (2007).
\end{flushleft}

\title{
  { \Large \bf \boldmath Search for the decay $B^+ \rightarrow K^+ \tau^\mp \mu^\pm$ }
}

\author{B.~Aubert}
\author{M.~Bona}
\author{D.~Boutigny}
\author{Y.~Karyotakis}
\author{J.~P.~Lees}
\author{V.~Poireau}
\author{X.~Prudent}
\author{V.~Tisserand}
\author{A.~Zghiche}
\affiliation{Laboratoire de Physique des Particules, IN2P3/CNRS et Universit\'e de Savoie, F-74941 Annecy-Le-Vieux, France }
\author{J.~Garra~Tico}
\author{E.~Grauges}
\affiliation{Universitat de Barcelona, Facultat de Fisica, Departament ECM, E-08028 Barcelona, Spain }
% ...
\collaboration{The \babar\ Collaboration}
\noaffiliation

\date{\today}

\begin{abstract}
   We present a search for the lepton flavor violating
   decay $B^+ \rightarrow K^+ \tau^\mp \mu^\pm$
   using 383 million $B\bar B$ events collected by the \babar\ experiment.
   The branching fraction for this decay can be substantially enhanced in
   new physics models.
      The kinematics of the tau from the signal
      $B$ decay are inferred from the $K^+$, $\mu$, and other $B$ in the event,
      which is fully reconstructed in one of a variety of hadronic decay
      modes, allowing the signal $B$ candidate to be fully reconstructed.
   We observe no excess of events over the expected background and
   set a limit of ${\cal B}(B^+ \rightarrow K^+ \tau \mu) < 7.7 \times 10^{-5}$ at 90\% confidence
   level, where the branching fraction is for the sum of the $K^+ \tau^- \mu^+$
   and $K^+ \tau^+ \mu^-$ final states.
   We use this result to improve a model-independent bound on the energy scale
   of flavor-changing new physics.
\end{abstract}

\pacs{
13.25.Hw, 
 14.40.Nd  
}

\maketitle

   Lepton flavor violation (LFV) thus far has only been observed in the
   neutrino sector
   \cite{superk,sno,kamland}.
   Manifestations of LFV in $B$ meson decays that have
   final states with charged leptons
   (e.g., $B\rightarrow K \ell \ell'$) are allowed in standard model
   interactions, if massive neutrinos are included \cite{SM}, but such
   processes occur only at the one-loop level and are extremely suppressed by
   powers of $m_\nu^2/M^2_W$.
   Branching fractions for lepton flavor violating $B$ decays can be
   substantially enhanced in many extensions of the standard
   model \cite{sher-and-yuan,black,he,fujihara}.
   The semileptonic decay $B \rightarrow K \tau \mu$ is likely to have
   higher sensitivity to new physics, when compared to leptonic decays
   such as $B^0 \rightarrow \tau \ell'$, since the latter is both helicity
   and
   Cabibbo-Kobayashi-Maskawa quark-mixing matrix
   (CKM) suppressed by a factor of $|V_{td}/V_{cb}|^2$.
   Some new physics models require flavor changing neutral currents (FCNC's)
   to occur at the one-loop level, as in the standard model.
   In other extensions, such as models with a $Z'$ or
   additional Higgs doublets, FCNC's occur naturally at the tree level,
   unless they are eliminated by imposing an {\it ad hoc} discrete symmetry.

   A limit on the process $B\rightarrow K\tau\mu$,
   which involves the second and third generations of both
   quarks and leptons, would provide a unique and powerful
   constraint on model parameters of grand unified theories.
   Cheng, Sher, and Yuan~\cite{cheng,sher-and-yuan} propose that
   in models with an extended Higgs sector the most
   natural value of the FCNC Yukawa couplings connecting
   generations $i$ and $j$ are proportional
   to $\sqrt{m_i m_j}/m_\tau$, which implies that FCNC's in
   these theories should be largest in processes involving the
   second and third generations.
   An observation of $B\rightarrow K\tau\mu$ would be an unambiguous
   sign of physics beyond the standard model.
   In this paper, we present the
   results of a search for $B\rightarrow K\tau\mu$.

   We use a data sample of 383 million $B \bar B$ pairs
   produced by the PEP-II asymmetric-energy $e^+e^-$ collider,
   running at the $\Upsilon(4S)$ resonance,
   collected by the \babar\ experiment~\cite{babar} at
   SLAC.
   Charged particles are identified using a Cerenkov radiation detector
   and $dE/dx$ measurements in the tracking system.
   Instrumentation embedded within the iron of the flux return for the
   1.5 T solenoid aids in the identification of muons.
   An electromagnetic CsI(Tl) crystal calorimeter (EMC) is used to
   reconstruct photons and identify electrons.

   The analysis strategy is to reconstruct the $\Upsilon(4S) \rightarrow B^+ B^-$
   in the search for $B^+ \rightarrow K^+ \tau \mu$~\cite{charge-conj}.
   One of the $B$ mesons ($B_{\rm tag}$) is fully reconstructed in one of
   a large number of hadronic final states, $B^- \rightarrow D^{(*)0} X^-$\cite{BpmSemiExcl}.
          The $X^-$ represents a system of charged and neutral hadrons with
          total charge $-1$ composed of $n_1 \, \pi^\pm$, $n_2 \, K^\pm$,
          $n_3 \, K^0_S$, and $n_4 \, \pi^0$, with $n_1+n_2\le 5$,
          $n_3 \le 2$, and $n_4 \le 2$.
          The $D^{*0}$ is reconstructed in the $D^0 \pi^0$ and $D^0\gamma$
          channels, the $D^0$ in the $K^-\pi^+$,
          $K^-\pi^+\pi^0$, $K^-\pi^+\pi^-\pi^+$, and $K_S^0 \pi^+\pi^-$ channels,
          and $K^0_S$ in the $\pi^+\pi^-$ channel.
   We search for the decay $B^+ \rightarrow K^+ \tau \mu$ using the
   remaining tracks in the event.
   The momentum vector of the signal $B$ candidate, $\vec p_{\rm sig}$, must be equal
   in magnitude and opposite in direction to that of $B_{\rm tag}$ in the
   center-of-mass (CM) frame.
   The $\tau$ candidate kinematic variables, $E_\tau$ and $\vec p_\tau$, are fully constrained
   by $\vec p_{\rm sig}$, the measured momenta of the $K^+$ and $\mu$ tracks, and
   the constraint $E_\tau = E_{\rm beam} - E_K - E_\mu$, where $E_{\rm beam}$ is the
   CM beam energy.
   The reconstructed $\tau$ invariant mass $m_\tau = \sqrt{E^2_\tau - p^2_\tau}$
   peaks sharply at the true $\tau$ mass for the signal.

   Events are required to contain a $B_{\rm tag}$ candidate with
   $m_{\rm ES} \equiv \sqrt{E^2_{\rm beam} - p_{\rm tag}^2} > 5.27$~GeV/c$^2$
   and $E_{\rm tag}$ consistent with $E_{\rm beam}$ within three standard deviations,
   where $p_{\rm tag}$ and $E_{\rm tag}$ are the momentum and energy of
   the $B_{\rm tag}$ candidate in the CM frame.
   A $B_{\rm tag}$ meson is fully reconstructed in about 0.2\% of Monte Carlo
   events where one $B^\pm$ decays to $K^\pm\tau\mu$.
   Even though the $\tau$ daughters are not needed to reconstruct $m_\tau$,
   we require the $\tau$ in the signal $B$ candidate to be consistent
   with a ``one-prong'' (i.e., one-charged-track) $\tau$ decay to reject combinatoric
   background.
   Therefore we require exactly three charged tracks in the event not associated
   with $B_{\rm tag}$ and with net charge opposite to that of $B_{\rm tag}$.
   Among these three tracks, we require: a kaon candidate with
   charge opposite $B_{\rm tag}$, a muon candidate, and a third track
   (the $\tau$ daughter) with charge opposite the muon candidate.
   The event is rejected if any of the three tracks is consistent with
   a proton hypothesis or if either of the two non-kaon tracks is consistent
   with a kaon hypothesis.
   Signal candidates are divided into three categories based on the
   properties of the $\tau$-daughter track: electron, muon, and pion.

   The kaon, muon, and electron particle identification criteria used
   in this analysis have momentum-dependent efficiencies and misidentification
   probabilities (fake rates).
   The kaon candidate must pass loose selection criteria, based on the
   measured Cerenkov angle and $dE/dx$ in the tracking system.
   Muon candidates, either from the $B$ decay or from the $\tau$
   decay, must pass minimum selection criteria that are
   85\% efficient for muons above 1.5~GeV/c and
   less than 10\% efficient for pions and kaons.
   Tau daughter electrons must pass minimum electron selection
   criteria that are 95\% efficient for electrons.
   More stringent electron and muon identification criteria
   for the tau daughter track are incorporated through
   a likelihood ratio described below.
   Tau daughters that do not pass either the electron
   or muon criteria fall into the pion category.

       For muon channel signal $B$ candidates, there are two muons in
       the final state: one from the $B$ decay (primary muon)
       and one from a leptonic $\tau$ decay.
       Of the two possible track assignments for the
       primary muon, we use the one that gives $m_\tau$ closest to the
       known $\tau$ mass.
     The bias in the background $m_\tau$ distribution
     for the muon channel
     from using this procedure is found to be negligible.

   Semileptonic $B$ decays can produce final states that appear identical
   to the signal.
   For example $B^+ \rightarrow \bar D^0 \mu^+ \nu_\mu$ followed by
   $\bar D^0 \rightarrow K^+ \pi^-$ produces a $K^+ \mu^+ \pi^- \nu$ final
   state.
   If the $\bar D^0$ decays semileptonically, the final state
   is $K^+ \mu^+ \ell^- \nu \bar\nu$.
   These backgrounds are easily removed by requiring that the invariant
   mass, $m(K\pi)$, of the kaon candidate and the oppositely-charged, signal-track
   candidate, when this track is assumed to be a pion, be
   greater than 1.95~GeV/c$^2$.
   This requirement is greater than 50\% efficient for the signal and removes
   about 99\% of the background from $B\bar B$ events.

\begin{figure}
     \begin{center}
         \includegraphics[width=\linewidth]{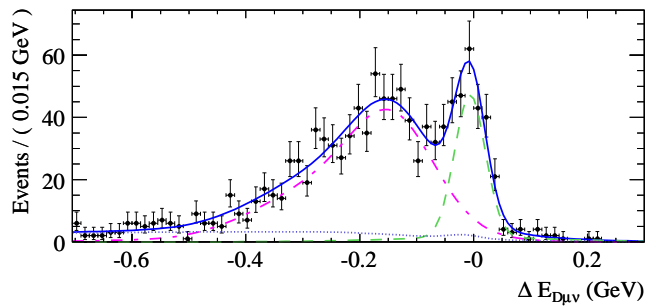}
       \caption{ Results of the $D\mu\nu$ control sample $\Delta E_{D\mu\nu}$ fit.
                 The points with error bars are the data,
                 the solid curve is the projection of the fit,
                 the dashed (dot-dashed) curve is the
                    $D^0\mu\nu$
                    ($D^{*0}\mu\nu$)
                    component,
                 and the dotted curve represents events from other sources.
                 }
       \label{fig:defit}
     \end{center}
   \end{figure}

   The $B^+ \rightarrow \bar D^{0} \mu^+ \nu_\mu$
   and $B^+ \rightarrow \bar D^{*0} \mu^+ \nu_\mu$, with $\bar D^0 \rightarrow K^+ \pi^-$,
   backgrounds also form the $D\mu\nu$ control sample, which we use to normalize the
   signal branching fraction.
   Events for the $D\mu\nu$ control sample are required to have $m(K\pi)$
   in the range [1.845, 1.885]~GeV/c$^2$ (within about three
   standard deviations
   of the $D^0$ mass).
   The neutrino momentum is calculated from $\vec p_{\rm tag}$ and the
   three tracks in the $D\mu\nu$ final state.
   We then compute
   \begin{equation}
      \Delta E_{D\mu\nu} = E_K + E_\mu + E_\pi + p_\nu - E_{\rm beam} = p_{\nu} - E_{\rm miss}.
   \end{equation}
   We use $\Delta E_{D\mu\nu}$ rather than $m_{\rm miss} = \sqrt{E_{\rm miss}^2 - p_{\nu}^2}$,
   similar to our $m_\tau$ reconstruction, because the expected $D^0\mu\nu$ missing
   mass is zero.
   The $\Delta E_{D\mu\nu}$ distribution for $D^0 \mu \nu$ decays is centered
   at zero, while for $D^{*0} \mu \nu$ events, it is shifted by $-150$~MeV and
   slightly asymmetric, due to the missing neutral particle from the
   $D^{*0} \rightarrow (\pi^0,\gamma) D^0$ decay.
   We determine the yield of $D^0\mu\nu$ and $D^{*0}\mu\nu$ events simultaneously
   in an unbinned maximum likelihood fit of $\Delta E_{D\mu\nu}$ (Figure~\ref{fig:defit}).

   For the $K^+ \tau \mu$ signal,
   a non-negligible source of $B\bar B$ background in the sample remaining
   after the $m(K\pi)$ requirement comes from $B^+ \rightarrow (c\bar c) K^+$ with
   $(c\bar c) \rightarrow \mu^+ \mu^-$, where $(c\bar c)$ is a charmonium
   resonance.
   This background mainly enters in the muon channel, but is also present
   in the pion channel.
   For the muon and pion channels this background
   is removed by requiring the invariant mass of the two non-kaon tracks,
   when both are assumed to be muons, to be outside of the ranges
   [3.03, 3.14]~GeV/c$^2$ and [3.60,3.75]~GeV/c$^2$,
   which are centered on the $J/\psi$ and $\psi(2S)$ resonances
   masses respectively.

   At this stage in the selection, the background is dominated by continuum events
   ($e^+e^- \rightarrow q\bar q$ with $q=u,d,s,c$).
   We suppress this background using a likelihood ratio
   \begin{equation}
      L_R = \frac{ \prod_i P_s(x_i) }
           { \prod_i P_s(x_i) + \prod_i P_b(x_i) }
   \end{equation}
   where $\vec x$ is a vector of four discriminating variables (see below)
   and $P_s(x_i)$ ($P_b(x_i)$) is the probability density function (PDF),
   which describes the signal (background) for variable $x_i$.
   The PDFs are separate for each of the three signal categories.
   The four discriminating variables are:
     $|\cos \theta_{\rm thr}|$ the magnitude of the cosine of the
     CM angle between the $B_{\rm tag}$ thrust axis and the thrust axis
     of the rest of the event;
     $\sum E_{\rm cal}$ the total neutral EMC energy that is not associated
     with $B_{\rm tag}$;
     the quality of the primary lepton identification;
     and the quality of the secondary lepton identification (for electron
     and muon channel candidates).
   We require a minimum $E_{\rm cal}$ energy of 50 MeV (100 MeV) for clusters
   in the barrel (forward endcap) to be included in $\sum E_{\rm cal}$.

   The $|\cos \theta_{\rm thr}|$ distribution is flat for signal and
   peaks near one for continuum.
   The $\sum E_{\rm cal}$ distribution peaks at zero for signal, while
   the background distribution is broad, peaking at around 1.5 GeV.
   The lepton quality is divided into four hierarchical, mutually-exclusive
   categories with fake rates decreasing with increasing quality rank.
   For the highest-quality muon candidate rank, the fake rates
   from pions and kaons are less than 2\%.
   The highest-quality electron candidate rank has a fake rate of less
   than 0.1\% for pions, as high as 3\% for low-momentum kaons, and
   below 0.4\% for kaons above 0.8~GeV/c.
   We fit signal and background Monte Carlo histograms of $|\cos \theta_{\rm thr}|$ 
   and $\sum E_{\rm cal}$ to define $P_s(x_i)$ and $P_b(x_i)$ for those
   variables.
   We use the relative fractions in the four lepton quality categories
   in the Monte Carlo samples for $P_s(x_i)$ and $P_b(x_i)$ for the primary and
   secondary lepton variables.

   We make a minimum $L_R$ requirement for each of the three signal
   categories (electron, muon, and pion)
   which has been optimized to give the lowest signal branching
   fraction limit under the assumption of no signal
   in the data.
   The signal region in $m_\tau$ is defined to be [1.65, 1.90]~GeV/c$^2$,
   which contains 90\% of the signal.
   The signal selection efficiency ($\epsilon_i$), including the $L_R$ requirements,
   in the $m_\tau$ signal region
   is 3.17\%, 2.04\%, and 2.13\% for the electron, muon, and pion
   channels respectively.
   The denominator of these $\epsilon_i$ is the same for all three
   and includes all tau decays.
   We have used a simple 3-body phase space decay model to generate
   our signal Monte Carlo sample.
   The systematic uncertainty on $\epsilon_i$ is determined by varying the
   signal and background PDFs for each $L_R$.
   Because the signal branching fraction is determined from the ratio of
   the signal and $D\mu\nu$ yields in the data, many systematic uncertainties associated
   with tracking, particle identification, and the $B_{\rm tag}$ reconstruction
   cancel.
   The amount of background, $b_i$, in the $m_\tau$ signal region is estimated
   from the number of events outside the $m_\tau$ signal region
   (the $m_\tau$ sidebands) in the ranges [0, 1.65] and [1.9, 3.5]~GeV/c$^2$ and
   the signal-to-sideband ratio from the background Monte Carlo sample.

   The signal branching fraction for each channel
   (${\cal B}_i$) is estimated using the relation
   \begin{equation}
   \label{eqn:sigb}
      {\cal B}_i = (n_i - b_i) / (\epsilon_i S_0),
   \end{equation}
   where $n_i$ is the observed number of events in the $m_\tau$
   signal region, $b_i$ is the expected background,
   $\epsilon_i$ is the signal efficiency,
   and $S_0$ is a common sensitivity factor given by
   \begin{equation}
      S_0 = \frac{N_{D\mu\nu}}{{\cal B}_{D\mu\nu}}
      \left( \frac{1}{\epsilon_{D\mu\nu}}\right)
      \left( \frac{ \epsilon^{K\tau\mu}_{\rm tag}}{\epsilon^{D\mu\nu}_{\rm tag}} \right),
   \end{equation}
   where $N_{D\mu\nu}$, ${\cal B}_{D\mu\nu}$, and $\epsilon_{D\mu\nu}$
   are the fitted yield, total branching fraction, and selection
   efficiency for the $D\mu\nu$ control sample and
   $\epsilon^{K\tau\mu}_{\rm tag}$ and
   $\epsilon^{D\mu\nu}_{\rm tag}$ are the $B_{\rm tag}$ efficiencies
   for the signal and $D\mu\nu$ samples respectively.
   The last factor $(\epsilon^{K\tau\mu}_{\rm tag}/\epsilon^{D\mu\nu}_{\rm tag})$
   is determined from the Monte Carlo samples and
   close to one $(0.922 \pm 0.052)$
   since
   the topology of the events in the signal and $D\mu\nu$ samples is
   very similar.
   We find $N_{D\mu\nu} = 867 \pm 52$ with $\epsilon_{D\mu\nu}=0.345\pm 0.008$
   and ${\cal B}_{D\mu\nu} = (3.29 \pm 0.22)\times 10^{-3}$~\cite{pdg},
   which gives $S_0 = (7.0 \pm 0.7)\times 10^5$.

   \begin{table*}[floatfix]
     \caption{ The number of events in the $m_\tau$ sidebands, $N_{\rm sb}$,
     for the Monte Carlo sample and the data;
     the ratio of background events, inside / outside the $m_\tau$
        signal region, BG ratio;
     the expected number of background events, $b_i$, and number of observed
        data events, $n_i$, in the $m_\tau$ signal region; and
     the signal selection efficiency $\epsilon_i$
     for each of the three channels.
     The first (second) uncertainty on $\epsilon_i$ is statistical (systematic).
     }
     \label{tab:results}
     \begin{tabular}{crccccc}
      \ Channel \  
      & \multicolumn{1}{c}{ \ \ $N_{\rm sb}$ (MC)  }
      &  \ \ $N_{\rm sb}$ (data) \ \
      &  \ \ \ BG ratio \ \ \ \
      &  \ \ \ \ \ $b_i$ \ \ \ \ \
      &  \ \ \ $n_i$ \ \ \
      &  \ \ \ \ $\epsilon_i$ (\%) \ \ \ \ \  \\
       \hline
       electron &       $5.2 \pm 1.3$  &        $5$            
                &   $0.10 \pm 0.05$    &   $0.5 \pm 0.3$
                &        1             &      $3.28 \pm 0.13 \pm 0.22$   \\
       muon     &       $0.7 \pm 0.5$  &        $2$            
                &   $0.30 \pm 0.15$    &   $0.6 \pm 0.3$
                &        0             &      $2.09 \pm 0.10 \pm 0.19$   \\
       pion     &      $6.9 \pm 1.6$  &       $14$            
                &   $0.13 \pm 0.04$    &   $1.8 \pm 0.6$
                &        2             &      $2.18 \pm 0.11 \pm 0.24$   \\
     \end{tabular}
   \end{table*}

\begin{figure}
     \begin{center}
         \includegraphics[width=\linewidth]{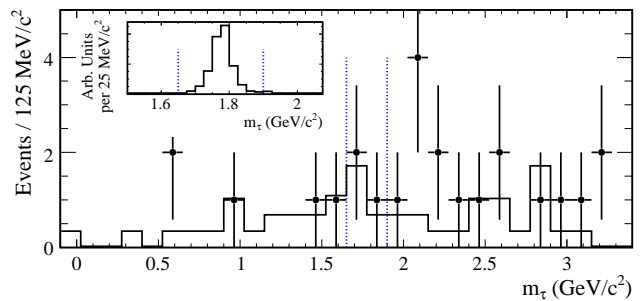}
       \caption{ Distributions of $m_\tau$
                 after all selection criteria have been applied
                 for the data (points with error bars),
                 background Monte Carlo sample (main histogram),
                 and signal Monte Carlo sample (inset histogram).
                 The dotted vertical lines show the $m_\tau$
                 signal region [1.65, 1.90] GeV/c$^2$.
                 }
       \label{fig:mtau}
     \end{center}
   \end{figure}

   The $m_\tau$ signal region in the data was kept blind during the development
   of the analysis, to avoid experimenter's bias.
   After all analysis decisions were made, we ``opened the box'' and found
   1, 0, and 2 events in the $m_\tau$ signal region for the
   electron, muon, and pion channels respectively.
   These totals are consistent with our expectations from background only,
   which are given in Table~\ref{tab:results}.
   Distributions of $m_\tau$ for the data and for the signal Monte Carlo sample
   are shown in Figure~\ref{fig:mtau}.
   The numbers of background events in the $m_\tau$ sidebands are consistent
   with our expectations from the Monte Carlo sample.

   The central value for the signal branching fraction is
   ${\cal B} = (0.8\ ^{+3.5}_{-2.3}) \times 10^{-5}$,
   where the uncertainties include both statistical and
   systematic sources.
   The three channels were combined by maximizing the likelihood, which
   is defined as the product of three Poisson probabilities in $n_i$, where
   the mean is given by Equation~\ref{eqn:sigb}.
   Uncertainties on the $b_i$, $\epsilon_i$, and $S_0$ parameters, which
   determine the Poisson mean for each channel, were included
   by convolving the Poisson PDFs with Gaussians in
   $b_i$, $\epsilon_i$, and $S_0$.
   The uncertainties on ${\cal B}$ correspond to the points where
   the log likelihood drops by 0.5 with respect to the maximum log
   likelihood value.
   We have verified, with a Monte Carlo study, that this maximum
   likelihood technique is unbiased and that the uncertainties are a
   reasonable estimate of one standard deviation.
   We find a 90\% confidence level upper limit on the signal branching
   fraction of ${\cal B} < 7.7 \times 10^{-5}$ using the prescription of
   Feldman and Cousins~\cite{fc} for defining the confidence belt.
         The central value and upper limit on the signal branching
         fraction are both limited by statistical uncertainties.

   In conclusion, we present the first search for the forbidden
   decay $B^+ \rightarrow K^+ \tau \mu$ using 383 million $B\bar B$
   pairs collected by the \babar\ experiment.
   The observed events are consistent with the background-only hypothesis.
   We set an upper limit of ${\cal B} < 7.7 \times 10^{-5}$ on the signal
   branching fraction at 90\% confidence level.
   This result can be used to improve the model-independent bound
   on the energy scale of new physics in flavor-changing operators
   reported in \cite{black} from $>2.6$~TeV to $>13$~TeV.

   \par
We are grateful for the excellent luminosity and machine conditions
provided by our \pep2\ colleagues, 
and for the substantial dedicated effort from
the computing organizations that support \babar.
The collaborating institutions wish to thank 
SLAC for its support and kind hospitality. 
This work is supported by
DOE
and NSF (USA),
NSERC (Canada),
CEA and
CNRS-IN2P3
(France),
BMBF and DFG
(Germany),
INFN (Italy),
FOM (The Netherlands),
NFR (Norway),
MIST (Russia),
MEC (Spain), and
STFC (United Kingdom). 
Individuals have received support from the
Marie Curie EIF (European Union) and
the A.~P.~Sloan Foundation.


\begin{thebibliography}{99}

   \bibitem{superk}  Y. Ashie {\it et al.}, The Super Kamiokande collaboration,
       Phys. Rev. {\bf D71}, 112005 (2005).
   \bibitem{sno}  S. N. Ahmed {\it et al.}, The SNO collaboration,
       Phys. Rev. Lett. {\bf 92}, 181301 (2004).
   \bibitem{kamland} T. Araki {\it et al.}, The KamLAND collaboration,
       Phys. Rev. Lett. {\bf 94}, 081801 (2005).
   \bibitem{SM} S. Weinberg, Phys. Rev. Lett., {\bf 19}, 1264 (1967);
               S. L. Glashow, J. Iliopoulos, and L. Maiani,
               Phys. Rev. {\bf D2}, 1285 (1970);  see also
               J. Erler and P. Langacker in \cite{pdg} and
               B. Kayser in \cite{pdg}.
   \bibitem{sher-and-yuan} M. Sher and Y. Yuan, Phys. Rev. {\bf D44}, 1461 (1991).
   \bibitem{black} D. Black, T. Han, H.-J. He, and M. Sher,
      Phys. Rev. {\bf D66}, 053002 (2002).
   \bibitem{he} X.-G. He, G. Valencia, and Y. Wang,
       Phys. Rev. {\bf D70}, 113011 (2004).
   \bibitem{fujihara} T. Fujihara {\it et al.},
       Phys. Rev. {\bf D73}, 074011 (2006).
   \bibitem{cheng} T. P. Cheng and M. Sher, Phys. Rev. {\bf D35}, 3484 (1987).
   \bibitem{babar} B. Aubert {\it et al.}, The \babar\ collaboration,
        Nucl. Instrum. Meth. {\bf A479}, 1 (2002).
   \bibitem{charge-conj}  By $B^+\rightarrow K^+ \tau \mu$ we mean
   either $B^+\rightarrow K^+ \tau^+ \mu^-$ or $B^+\rightarrow K^+ \tau^- \mu^+$;
   also, charge-conjugate final states are included unless
     explicitly stated otherwise.
   \bibitem{BpmSemiExcl} B. Aubert {\it et al.}, The \babar\ collaboration,
       Phys. Rev. {\bf D74}, 031103 (2006).
   \bibitem{pdg} W.-M. Yao {\it et al.}, The Particle Data Group,
        J. Phys. {\bf G33}, 1 (2006).
   \bibitem{fc} G. J. Feldman and R. D. Cousins,
       Phys. Rev. {\bf D57}, 3873 (1998).
   \end{thebibliography}
\end{document}